\documentclass{nature}
\usepackage{epsfig}
\usepackage{color}
\usepackage{amsmath}
\usepackage{lscape}
\usepackage{amssymb}

\begin{document}

\title{Crystallized and amorphous vortices in rotating atomic-molecular Bose-Einstein condensates}
\author{Chao-Fei Liu$^{1,2}$, Heng Fan$^{1}$, Shih-Chuan Gou$^{3}$, and Wu-Ming Liu$^{1\star}$}

\maketitle

\begin{affiliations}
\item
Beijing National Laboratory for Condensed Matter Physics,
Institute of Physics, Chinese Academy of Sciences,
Beijing 100190, China

\item
School of Science, Jiangxi University of Science and Technology, Ganzhou 341000, China

\item
Department of Physics, National Changhua University of Education, Changhua 50058, Taiwan


$^\star$e-mail: wliu@iphy.ac.cn

\end{affiliations}

\begin{abstract}

Vortex is a topological defect with a quantized winding number of the phase in superfluids and superconductors. Here, we investigate the crystallized (triangular, square, honeycomb) and amorphous vortices in rotating atomic-molecular Bose-Einstein condensates (BECs) by using the damped projected Gross-Pitaevskii equation. The amorphous vortices are the result of the considerable deviation induced by the interaction of atomic-molecular vortices. By changing the atom-molecule interaction from attractive to repulsive, the configuration of vortices can change from an overlapped atomic-molecular vortices to carbon-dioxide-type ones, then to atomic vortices with interstitial molecular vortices, and finally into independent separated ones. The Raman detuning can tune the ratio of the atomic vortex to the molecular vortex. We provide a phase diagram of vortices in rotating atomic-molecular BECs as a function of Raman detuning and the strength of atom-molecule interaction.

\end{abstract}


The realization of Bose-Einstein condensate (BEC) in dilute atomic gas is one of the greatest achievements for observing the intriguing quantum phenomena on the macroscopic scale. For example, this system is very suitable for observing the quantized vortex \cite{Madison}, and the crystallized quantized vortex lattice \cite{AboShaeer, Engels}. Furthermore, it is found that vortex lattices in rotating single atomic BEC with dipole interaction can display the triangular, square, ``stripe", and ``bubble" phases \cite{Cooper}. In two-component atomic BEC, the vortex states of square, triangular, double-core and serpentine lattices are showed according to the intercomponent coupling constant and the geometry of trap \cite{Kasamatsu}. Considered two components with unequal atomic masses and attractive intercomponent interaction, the exotic lattices such as two superposed triangular, square lattices and two crossing square lattices tilted by $\pi/4$ are indicated \cite{Kuopanportti}. Generally speaking, the crystallization of vortices into regular structures is common in the single BEC and the miscible multicomponent BECs under a normal harmonic trap. Vortices in atomic BECs have attracted much attentions \cite{Wucj, Malomeda, Malomedb, Jiac, Wangds, Hanwei, liuDPGPE, CFLiu, liucf, Su}. However, it is not very clear the crystallization of vortices in atomic-molecular BECs \cite{Alexander, Woo, Timmermans, Wynar, Donley, McKenzie, Cusack, Basu, Zhou, Ling, Linghy, Xu, Heinzen, Drummond, Abdullaev, Cruz, Gupta}.

The molecular BEC can be created by the magnetoassociation (Feshbach resonance) of cold atoms to molecules \cite{ Donley}, and by the Raman photoassociation of atoms in a condensate \cite{Wynar, McKenzie}. The atomic-molecular BEC provides a new platform for exploring novel vortex phenomena. It is shown recently that the coherent coupling can render a pairing of atomic and molecular vortices into a composite structure that resembles a carbon dioxide molecule \cite{Woo}. Considering both attractive and repulsive atom-molecule interaction, Woo \emph{et al.} have explored the structural phase transition of atomic-molecular vortex lattices by increasing the rotating frequency. They observed the Archimedean lattice of vortex with the repulsive atom-molecule interaction. In fact, atom-molecule interaction can be either attractive or repulsive with large amplitude by using the Feshbach resonance \cite{ Donley, Gupta}. In addition, we know that the population of atom and molecule in atomic-molecular BECs can be tuned by the Raman photoassociation \cite{Ling, Heinzen, Drummond, Abdullaev, Cruz, Gupta}. Then, we may wonder whether the combination control of Raman detuning and atom-molecule interaction may induce nontrivial vortex states and novel vortex phenomena. This seems not be well explored, especially in the grand canonical ensemble \cite{Herzog, Kocharovsky, Cockburn}. Furthermore, similarly to the normal system of two-component BECs \cite{ Kasamatsu}, a phase diagram of vortices in rotating atomic-molecular BECs is required to provide a full realization of the nontrivial vortex phenomenon.

In this report, we study the crystallized and amorphous vortices in rotating atomic-molecular BECs \cite{Timmermans, Alexander, Wynar, Donley, McKenzie, Cusack, Basu, Zhou, Ling, Linghy, Xu, Heinzen, Drummond, Abdullaev, Cruz, Gupta}. Amorphous vortices are the result of the considerable deviation induced by the interaction of atomic-molecular vortices. The phase diagram indicates that atom-molecule interaction can control the atomic-molecular vortices to suffer a dramatic dissociation transition from an overlapped atomic-molecular vortices with interlaced molecular vortices to the carbon-dioxide-type atomic-molecular vortices, then to the atomic vortices with interstitial molecular vortices, and finally to the completely separated atomic-molecular vortices. This result is in accordance with the predicted dissociation of the composite vortex lattice in the flux-flow of two-band superconductors \cite{Lin}. The Raman detuning adjusts the population of atomic-molecular BECs and the corresponding vortices. This leads to the imbalance transition among vortex states. This study shows a full picture about the vortex state in rotating atomic-molecular BECs.

\section*{Results}
\subsection{The coupled Gross-Pitaevskii equations for characterizing atomic-molecular Bose-Einstein condensates.}
We ignore the molecular spontaneous emission and the light shift effect \cite{ McKenzie, Heinzen, Drummond, Gupta}.
According to the mean-field theory, the coupled equations of atomic-molecular BEC \cite{ Gupta, Woo, Tikhonenkov} can be written as
\begin{eqnarray}\label{GP2}
i\hbar\frac{\partial\Psi_{a}}{\partial t}=[-\frac{\hbar^{2}\nabla^{2}}{2M_{a}}+\frac{M_{a}\omega^{2}(x^{2}+y^{2})}{2}]\Psi_{a}-\Omega\widehat{L}_{z}\Psi_{a} \notag \\
+(g_{a}|\Psi_{a}|^{2}+g_{am}|\Psi_{m}|^{2})\Psi_{a} +\sqrt{2}\chi \Psi_{a}^{*}\Psi_{m}, \notag\\
i\hbar\frac{\partial\Psi_{m}}{\partial t}=[-\frac{\hbar^{2}\nabla^{2}}{2M_{m}}+\frac{M_{m}\omega^{2}(x^{2}+y^{2})}{2}]\Psi_{m}-\Omega\widehat{L}_{z}\Psi_{m} \notag \\
+(g_{am}|\Psi_{a}|^{2}+g_{m}|\Psi_{m}|^{2})\Psi_{m} +\frac{\chi}{\sqrt{2}} \Psi_{a}^{2}+\varepsilon\Psi_{m},
\end{eqnarray}
where $\Psi_{j}(j=a,m)$ denotes the macroscopic wave
function of atomic condensate and molecular condensate respectively, the coupling constants are,
$g_{a}=\frac{4\pi\hbar^{2}a_{a}}{M_{a}}$, $g_{m}=\frac{4\pi\hbar^{2}a_{m}}{M_{m}}$, and $g_{am}=\frac{2\pi\hbar^{2}a_{am}}{M_{m}M_{a}/(M_{m}+M_{a})}$,
also $M_{a}$ ($M_{m}$) is the mass of atom (molecule), $\omega$ is the trapped frequency,
$\Omega$ is the rotation
frequency, $\widehat{L}_{z}$ [$\widehat{L}_{z}=-i\hbar(x\partial_{y}-y\partial_{x})$] is the $z$ component of the
orbital angular momentum. The parameter $\chi$ describes the conversions of atoms into molecules due to stimulated Raman transitions.
$\varepsilon$ is a parameter to characterize Raman detuning for a two photon resonance \cite{Wynar, McKenzie, Heinzen, Drummond, Gupta}.

In real experiment, it is observed that the coherent free-bound stimulated Raman transition can cause atomic BEC of $^{87}$Rb to generate a molecular BEC of $^{87}$Rb \cite{Wynar}.
In numerical simulations, we use the parameters of atomic-molecular BECs of $^{87}$Rb system
with $M_{m}=2M_{a}=2m$ ($m=144.42\times 10^{-27}Kg$), $g_{m}=2g_{a}$ ($a_{a}=101.8a_{B}$, where $a_{B}$ is the Bohr radius), $\chi=2\times10^{-3}$, and the trapped frequency $\omega=100\times2\pi$. Note that if the change in energy in converting two atoms into one molecule ($\Delta U=2U_{Ta}-U_{Tm}$) \cite{Wynar}, not including internal energy, approaches zero, we can obtain the value $2g_{a}=g_{m}$.
The unit of length, time, and energy correspond to $\sqrt{\hbar/(m\omega)}$ ($\approx1.07\mu m$), $\omega^{-1}$ ($\approx1.6 \times 10^{-3}s$), and $\hbar\omega$, respectively.


\subsection{Crystallized and amorphous vortices in rotating atomic-molecular Bose-Einstein condensates.}

With rotation frequency $\Omega=0.8\omega$, we study the influence of atom-molecule interaction on the formation of vortices. Figure 1 displays the densities and phases obtained under the equilibrium state with different atom-molecule interactions. The first and the second columns are the densities of the atomic BEC and the molecular BEC, respectively. The third column is the total density. The fourth and the fifth column are the corresponding phases of atomic and molecular BECs, respectively. The vortices can be identified in the phase image of BECs.

The composite of atomic and molecular vortices locates at the trap center to lower the system's energy. For the case of attractive atom-molecule interaction ($g_{am}=-0.87g_{a}$), vortices form a square lattice [see Fig. 1(a)]. Interestingly, we can observe that the size of the molecular vortices can be divided into two types: one is big, and the other is small. Each atomic vortex has approximately the same size. Figure 2(a) further plots the vortices, where the atomic vortices overlap with a molecular one. The overlapping of atomic and molecular vortices causes the size of molecular vortices to become big. Thus, we obtain different size of molecular vortices in the same experiment.

It is easy to understand the size enlargement of molecular vortices which are overlapping with atomic ones.
The size of vortex reflects the healing length [$\xi=\hbar(2mg\overline{n})^{-1/2}$, where $\overline{n}$ is the uniform
density in a nonrotating cloud \cite{Fischer}] of the BEC because within this distance, the order parameter `heals' from zero up to its bulk value.
The attractive interspecies interaction implies that the densities of the two BECs would have a similar trend to decrease and increase.
It also causes some molecular vortices to overlap with atomic ones.
In addition, the density of atomic BEC forms local nonzero minima at the region of the left molecular vortex [see Fig. 2(a)].
Here, the size of atomic vortices is obvious bigger than that of molecular vortices.
Therefore, the local density of molecular vortices follows that of atomic vortices and the size becomes big when molecular vortices overlap with atomic vortices.

When $g_{am}=0$, atomic vortex lattices are triangular and the molecular vortices are amorphous state [see Fig. 1(b) and Fig. 2(d)]. Meanwhile, the total density (the third column) indicates that the molecular vortices and the atomic vortices form some structure like the carbon dioxide, which is also observed by a different method \cite{Woo}. Figure 2(b) shows an enlarged configuration of the carbon dioxide vortices. Here, the size of atomic vortices is much larger than that of the molecular one. With repulsive interaction ($g_{am}=0.87g_{a}$), vortex lattices are approximately hexagonal with a little deviation [see Fig. 1(c)]. Increasing atom-molecule interaction up to $g_{am}=2g_{a}$ [see Fig. 1(d)], atomic-molecular BECs separate into two parts, molecular BEC locating at the center and atomic BEC rounding it. The results are understandable, since the mass of a molecule is twice as that of atom, molecular BEC tends to locate at the center. This is much different from that of the normal two-component BECs, where the same mass and intraspecies interactions are considered \cite{Kasamatsu}.

Figures 2(c)-2(f) further illuminate the position of vortices. Note that we do not point out the vortices where the densities of BECs are very low. We approximately view the vortex lattice as triangular, square, etc, although some vortices may deviate from the regular lattice slightly. The distance of adjacent lattice sites of atomic vortices is $\sqrt{2}$ times of that of molecular vortices [see Fig. 2(c)]. We have plotted a green circle to differentiate these vortices as two parts. The atomic vortices construct an approximately quadrangle lattice, especially near the center region, the atomic vortices overlap with a molecular one locating among four adjacent molecular vortices. Thus, vortex position indicates that vortices density of atomic BEC is half of that of molecular vortices. The atomic vortices expand over to the outskirts of the lattice where no overlapped molecular vortices appear [see Fig. 2(c)].

Figure 2(d) indicates that the carbon dioxide structure is not fixed in the same orientation. Similarly to Fig. 2(c), the carbon dioxide structure only exists at the center. However, the deviation of molecular vortices from the red lines d, e, and f is so large that we have to view the molecular vortices as an amorphous state. Vortex position in Fig. 2(e) shows that atomic vortices form the triangle lattice. All molecular vortices are distributed among atomic vortices, forming the hexagonal lattices without overlapping. Certainly, atomic vortices and molecular vortices are separated in Fig. 2(f) according to the immiscibility of atomic-molecular BECs with strong $g_{am}$. We can conclude that the strength of atom-molecule interaction can adjust the composite degrees of vortices, and cause the overlapping composite, carbon-dioxide-type composite, interstitial composite and separation.

Furthermore, we find that the lattice configuration of vortices is very complex when atomic vortices and interstitial molecular vortices coexist. In Fig. 1(c), atomic vortices form the triangular lattice and interstitial molecular vortices display the honeycomb lattice. We further plot the densities of atomic BEC and molecular BEC at various cases in Fig. 3. When the number of atoms is much more than that of molecules, vortices in atomic BEC tend to form the triangular lattice, and vice versa. The lattice configurations are triangular in Figs. 3(a2), (b2), (e1) and (f1). Atomic vortices display square lattice in Figs. 3(a1)-3(c1). In all other subplots, the lattices are irregular and can be viewed as the amorphous state. For example, the number of adjacent molecular vortices which form bubbles \cite{Cooper} around some atomic vortices is not six but five in Figs. 3(d2)-(f2). In fact, the regular structures imply that both long-range order and short-range order should be remained. Thus, the observed random configuration is really amorphous.


\subsection{The phase diagram of rotating atomic-molecular Bose-Einstein condensates.}

To explore the phase diagram of atomic-molecular vortices, we firstly show the modulation effect of Raman detuning for the number of vortices in rotating atomic-molecular BECs. Figures 4(a)-(d) show the relationship between vortices number and Raman detuning. Generally speaking, the number of molecular vortices decreases monotonously as Raman detuning increases. As we can see for $g_{am}=-0.87g_{a}$, 0, $0.87g_{a}$ and $2g_{a}$, the slope of the number of molecular vortices is $-2.7$, $-3.4$, $-4.2$ and $-7.8$, respectively. Increasing of the strength of atom-molecule interaction, the faster the number of molecular vortices decreases as Raman detuning increasing. The number of atomic vortices approaches to 40 as the Raman detuning increasing. Thus, the ratio of atomic vortices and molecular vortices is not fixed as the Raman detuning changes in atomic-molecular BECs. Furthermore, we calculate the number of composite vortices, i.e., the atom-vortex number $C'_{a}$ and the molecule-vortex number $C'_{m}$ in the green circles in Figs. 2(c)-(e), and define the parameter $P_{m}=100 C'_{a}/C'_{m}$. $P_{a}$ in Figs. 4(a), (b), and (c) are almost around the value of 50, i.e, $C'_{a}:C'_{m}\approx 1:2$. Thus, the composite vortices keep the ratio 1:2 approximately. The deviation of $P_{a}$ from dash black line with the value of 50 mainly comes from vortices at the boundary. Certainly, vortex number in pure atomic BEC is independent of both atom-molecule interaction and Raman detuning.

Now, we further indicate the modulation effect of Raman detuning for particle numbers of atomic-molecular BEC of $^{87}$Rb [see Figs. 4(e)-(h)].
The particle numbers in equilibrium state depend on the system itself. For attractive atom-molecule interaction $g_{am}=-0.87g_{a}$, both of atom number and molecule number decrease when the Raman detuning increases. Meanwhile, atom number always is greater than molecule number [see Fig. 4(e)]. For the limit case of $g_{am}=0$, atom number keeps unchanged and only molecule number decreases as the Raman detuning increases [see Fig. 4(f)]. When the interaction is repulsive ($g_{am}>0$), molecule number keeps decreasing but atom number increases [see Figs. 4(g) and (h)]. When the repulsive interaction is up to $g_{am}=2g_{a}$, the single molecular BEC or the single atomic BEC can be obtained by adjusting Raman detuning from $-4\hbar\omega$ to $14\hbar\omega$.  The particle numbers can characterize the possible regions for the existence of atomic-molecular vortices.


Figure 5 plots the phase diagram of atomic-molecular BECs. The stable atomic-molecular BECs system exists only when atom-molecule interaction is larger than $-\sqrt{g_{a}g_{m}}$. When the Raman detuning is large enough, single atomic BEC occurs. Oppositely, if the Raman detuning is low enough, production changes into pure molecular BEC. Between these two regions, it is atomic-molecular BECs, where AMBEC(I) denotes the miscible mixture and AMBEC(II) stands for the phase separated mixture. Therefore, to explore atomic-molecular vortices, we mainly focus on AMBEC(I) region.

According to above analysis about atomic-molecular vortices and the corresponding atomic-molecular BECs, we calculate lots of other results and finally give a vortex phase diagram to summarize the vortex structures in Fig. 5. In (1) region [$-\sqrt{g_{a}g_{m}}<g_{am}<-0.1g_{a}$], atomic-molecular vortices form the square lattice where the overlapped atomic-molecular vortices and the molecular vortices interlacedly exist. The carbon-dioxide-type atomic-molecular vortices occur in (2) region [$-0.1g_{a}\leq g_{am}\leq 0.4g_{a}$]. Atomic vortices with interstitial molecular vortices emerge in region (3). In the AMBEC(II) region, atomic vortices and molecular vortices are separated. Certainly, in the region of atomic BEC (molecular BEC), atomic vortices (molecular vortices) favor to form the triangular lattice. The green region indicates that the created atomic and molecular vortices fully match with each other by roughly the ratio $1:2$. Above the green region, more atomic vortices occur. Below the green region, more molecular vortices appear. Table I shows a summary of the details of various vortices in rotating atomic-molecular BECs.

Interestingly, the vortex phase diagram indicates some exotic transitions. (i) Imbalance transition: The increase of Raman detuning causes more atomic BEC. Thus, the pure molecular vortices change into carbon-dioxide-type atomic-molecular vortices (atom vortices with interstitial molecular vortices, and separated atomic-molecular vortices), and finally into single atomic vortices in region of $0< g_{am}\leq 0.5g_{a}$ ($0.5g_{a}<g_{am}\leq \sqrt{g_{a}g_{m}}$, and $g_{am}>\sqrt{g_{a}g_{m}}$, respectively). In the region of $-\sqrt{g_{a}g_{m}}<g_{am}<-0.1g_{a}$ ($-0.1g_{a}\leq g_{am}\leq 0$), the interlaced-overlapped atomic-molecular vortices (the carbon-dioxide-type atomic-molecular vortices) become into atomic vortices under a very high detuning parameter. (ii) Dissociation transition: By changing the atom-molecule interaction from attractive to repulsive, the composite atomic-molecular vortices change from overlapped to carbon-dioxide-type and finally into the independent separated ones.

\section*{Discussion}

In this report, we focus on the strength of atom-molecule interaction and the Raman detuning term. The form of Hamiltonian in this paper is like that in the Ref. \cite{Woo}. In fact, a real experiment would include lots of other factors such as the light shift effect \cite{ McKenzie, Heinzen, Drummond, Gupta}, decay due to spontaneous emission \cite{Heinzen}. In Ref. \cite{Gupta}, Gupta and Dastidar have considered a more complicated model when they study the dynamics of atomic and molecular BECs of $^{87}$Rb in a spherically symmetric trap coupled by stimulated Raman photoassociation process. In fact, the light shift effect almost has the same function as the Raman detuning term. Thus, it can be contributed to the Raman detuning term. This is the reason why we do not consider the light shift term in Hamiltonian like that in Ref. \cite{Gupta}, but follows the form in Ref. \cite{Woo}.

In real experiment, it is believed that the single molecular BEC would occur when the Raman detuning goes to zero \cite{ Heinzen , McKenzie }. However, the measure of the remaining fraction of atom does not reach the minimum when Raman detuning is zero \cite{ McKenzie}. With the adiabatic consideration, the dynamical study also agrees with this point \cite{Gupta}. In fact, they show the evolutionary process of creating a molecular BEC from a single atomic BEC. Thus, particle number of molecular BEC varies with time but not fixed. The resonance coupling would cause the atomic BEC to convert into a molecular one as much as possible, but the molecular BEC also will convert into the atomic one. Therefore, the results in Ref. \cite{ McKenzie, Gupta} only shows a temporary conversion of atoms into molecules. In fact, when we use single atomic BEC as the initial condition and set $\frac{\gamma_{j}}{k_{B}T}=0$, the temporary conversion of atomic BEC into molecular BEC can be observed with current damped projected Gross-Pitaevskii equations.

It is obvious that the Raman detuning term in the Hamiltonian behaves just like the chemical potential to control the system's energy.
The external potential for atomic BEC is fixed to be $V_{a}(r)$ and molecular BEC experiences the trap potential $V_{m}(r)+\varepsilon$.
Here, our method initially derives from the finite-temperature consideration:
the system is divided into the coherent region with the energies of the state below $E_{R}$
and the noncoherent region with the energies of the state above $E_{R}$ \cite{ Rooney, Bradley}.
So, our method will behavior just likes to catch the particles with a shallow trap and exchange particles with an external thermal reservoir.
But ultimately we remove the external thermal reservoir to get system to the ground state.
Raman detuning changes the depth of shallow trap to $\mu_{m}-\varepsilon$.
The molecular BEC will be converted by atoms until the system reaches the equilibrium state.
Therefore, a maximum of creating molecular BEC does not occur at the equilibrium state when Raman detuning varies.
Instead, molecule number decreases monotonously when Raman detuning increases.


Why do atomic-molecular vortices display so rich lattice configurations?
In fact, atomic vortices and molecular vortices tend to be attractive in region (1) and (2).
Otherwise, the overlapped atomic-molecular vortices and the carbon-dioxide-type ones can not occur.
The attractive force makes atomic vortices and molecular vortices behave similarly.
Thus, both atomic and molecular vortex lattices in region (1) are square.
In region (2), atomic vortices display the triangular lattice.
Molecular vortices seem to follow the triangular lattice but the interaction among vortices causes the considerable deviation.
Obviously, the $CO_2$-type structures do not follow the fixed direction, i.e., long-range order vanishes but there is still short-range order.
Thus, we have to view molecular vortices as the amorphous state.
In region (3), atomic vortices and molecular vortices can not form the carbon dioxide structure.
Because the size of molecular vortices is smaller than that of atomic vortices, it tends to locate at the interval of the lattice of atomic vortices.
When the number of one component is much more than that of the other, the vortices of this component dominate over the vortices of the other component.
The former is easy to form the regular vortex lattice. The latter has to follow the interaction of the former and forms the vortex lattice.
The amorphous state originates from the competition between atomic vortices and molecular vortices,
especially when the number of atom and molecule has the considerable proportion [see Figs. 3(d1) and 3(d2)].
In that case, short-range order is only partly kept and ultimately long-range order is destroyed.
Certainly, this also causes the distribution of vortices in one component is relatively regular and that in the other component is amorphous.

The structural phase transitions of vortex lattices are explored through tuning the atom-molecule coupling coefficient and the rotational frequency of the system \cite{Woo}. Certainly, the Archimedean lattice of vortices in Ref. \cite{Woo} is one of the interstitial-composite-structures. Here, we show the crystallized and amorphous vortices by the combined control of Raman detuning and atom-molecule interaction. In fact, when we increase the value of $\chi$, the $CO_2$-type structure of vortices are easy to be created. Even the interstitial-composite structure we now obtain in Fig. 3 would transfer into the $CO_2$-type structure if $\chi$ is big enough. We have also considered the effect of rotation frequency. With the attractive interaction of atom-molecule ($g_{am}=-0.87g_{a}$), Figure 6 shows various rotation frequencies to produce the vortices. Figure 6(a) indicates that no vortex would occur with $\Omega=0$. For $\Omega=0.2\omega$, only one molecular vortex is induced. In atomic BEC, the phase indicates no vortex is created although there is a local minimum of density near the center. For $\Omega=0.4\omega$, the phase indicates that there is an atomic vortex. In fact, we find the atomic vortex is overlapped with a molecular vortex. Undoubtedly, more and more vortices emerge when rotation frequency increases. When the rotation frequency is up to $\Omega=0.8\omega$, we can obtain a regular square vortex lattice. Meanwhile, each atomic vortex is overlapped with a corresponding molecular vortex. Obviously, vortices and vortex lattice may not be induced with a slow rotation. This is the reason why we favor to investigate the vortices with a fast rotation in Figs. 1-4.

We now show that ultracold Bose gases of $^{87}$Rb atoms are a candidate for observing the predicted atomic-molecular vortices. By initially using a large atomic BEC of $^{87}$Rb (the atom number is up to $3.6\times10^{5}$ in Wynar's experiment \cite{Wynar}), the Raman photoassociation of atoms \cite{Heinzen, Wynar, McKenzie, Drummo, Hope} can produce the corresponding molecular BEC with partial of the atoms. By loading a pancakelike optical trap $V_{j=a, m}(x,y,z)=\frac{M_{j}[\omega^{2}(x^{2}+y^{2})+\omega_{z}z^{2}]}{2}$, with trapping frequencies $\omega_{z}\gg\omega$ \cite{Madison, AboShaeer, Engels}, the 2D atomic-molecular BECs may be prepared. It is convenient to use the laser to rotate the atomic-molecular BECs and induce the atomic-molecular vortices. Meanwhile, the whole system should be further quenched to a lower temperature to approach the ground state by the evaporative cooling techniques. The resulting atomic-molecular vortices may be visualized by using the scanning probe imaging techniques. All the techniques are therefore within the reach of current experiments.

In summary, we have observed various new atomic-molecular vortices and the lattices controlled by atom-molecule interaction and Raman detuning. Including the regular vortex lattices, we have displayed amorphous vortex state where vortices do not arrange regularly but like amorphous materials. We have obtained the vortex phase diagram as function of Raman detuning and atom-molecule interaction in the equilibrium state. Vortex configuration in atomic-molecular BECs includes the overlapped atomic-molecular vortices, the carbon-dioxide-type vortices, the atomic vortices with interstitial molecular vortices, and the completely separated atomic-molecular vortices. The lattice configuration of vortex mainly depends on atom-molecule interaction. For example, the overlapped atomic-molecular vortices display the square lattice. When the carbon-dioxide-type vortices occur, atomic vortices show the triangular lattice and molecular vortices show the amorphous state. Atomic vortices and interstitial molecular vortices can show several types of lattice, such as triangular, honeycomb, square and amorphous. And both atomic and molecular vortices show the triangular lattice in the incomposite region and in single BEC. Our results indicate that atom-molecule interaction can control the composite of atomic and molecular vortices, and can also cause novel dissociation transition of vortex state. Furthermore, the Raman detuning can control the numbers of particles in atomic-molecular BECs and approximately lead to the linear decrease of molecular vortices. This may induce the imbalance transition from atomic-molecular vortices to pure atomic (molecular) vortices. This study shows rich vortex states and exotic transitions in rotating atomic-molecular BECs.

\section*{Methods}

We use the damped projected Gross-Pitaevskii equation (PGPE) \cite{ Rooney} to obtain the ground state of atomic-molecular BEC.
By neglecting the noise term according to the corresponding stochastic PGPE \cite{ Bradley}, the damped PGPE is described as
\begin{equation}
d\Psi_{j}=\mathcal{P}\{-\frac{i}{\hbar}\widehat{H}_{j}\Psi_{j}dt+\frac{\gamma_{j}}{k_{B}T}(\mu_{j}-\widehat{H}_{j})\Psi_{j}dt\},
\end{equation}
where, $\widehat{H}_{j}\Psi_{j}=i\hbar\frac{\partial\Psi_{j}}{\partial t}$, $T$ is the final temperature, $k_{B}$ is the Boltzmann constant, $\mu_{j}$ is the chemical potential, and $\gamma_{j}$ is the growth rate for the $j$th component. The projection operator $\mathcal{P}$ is used to restrict the dynamics of atomic-molecular BEC in the coherent region. Meanwhile, we set the parameter $\frac{\gamma_{j}}{k_{B}T}=0.03$.
The initial state of each $\Psi_{j}$ is generated by sampling the grand canonical
ensemble for a free ideal Bose gas with the chemical potential $\mu_{m,0}=2\mu_{a,0}=8\hbar\omega$. The final chemical potential of the noncondensate band are altered to the values $\mu_{m}=2\mu_{a}=28\hbar\omega$.



\begin{addendum}

\item [Acknowledgement]

C. F. L. was supported by the NSFC under Grant No. 11247206, No. 11304130, No. 11365010 and the Science
and Technology Project of Jiangxi Province, China (Grant No. GJJ13382).
S.-C. G. was supported by the National Science Council,
Taiwan, under Grant No. 100-2112-M-018-001-MY3, and
partly by the National Center of Theoretical Science.
W. M. L. is supported
by the NKBRSFC under Grants No. 2011CB921502,
No. 2012CB821305, the NSFC under Grants
No. 61227902, No. 61378017, and No.11311120053. 


\item [Author Contributions]

W.M.L. conceived the idea and supervised the overall research. C.F.L. and. S.C.G. designed and
performed the numerical experiments. C.F.L. and H.F. wrote the
paper with helps from all other co-authors.

\item [Competing Interests]
The authors declare that they have no competing financial interests.

\item [Correspondence]
Correspondence and requests for materials should be addressed to Liu, Wu-Ming.
\end{addendum}

\clearpage

\newpage

\textbf{Figure 1 The densities and phases of atomic-molecular BECs of $^{87}$Rb when the system reaches the equilibrium state.} The rotation frequency is $\Omega=0.8\omega$. The strength of atom-atom interaction is $g_{a}$ with the scattering length $a_{a}=101.8a_{B}$. The strength of molecule-molecule interaction $g_{m}$ is twice as much as that of atom-atom interaction. The strength of atom-molecule interaction and Raman detuning is set as (a) $g_{am}=-0.87g_{a}$, $\varepsilon=14\hbar\omega$, (b) $g_{am}=0$, $\varepsilon=14\hbar\omega$, (c) $g_{am}=0.87g_{a}$, $\varepsilon=14\hbar\omega$ and (d) $g_{am}=2g_{a}$, $\varepsilon=7\hbar\omega$. Note that the fourth and fifth columns are the phases of atomic and molecular BECs, respectively.
The unit of length is $1.07\mu m$.

\bigskip

\textbf{Figure 2 Vortex configurations and vortex position.}
(a) The scheme of composite vortices in Fig. 1(a). The size of the right molecular vortex which overlaps with an atomic vortex is bigger than the left one. (b) The scheme of carbon-dioxide-type vortex structure in Fig. 1(b). The size of the atomic vortex is bigger than that of the molecular vortex. The red, black and blue indicate the densities of atomic BEC, molecular BEC and the sum, respectively.
(c), (d), (e) and (f) show the position of vortices in Figs. 1(a)-1(d), respectively.
The circle ($\circ$) and asterisk ($\ast$) are the position of vortices formed by atomic BEC and molecular BEC, respectively.
In (c), the red lines indicate that vortices can array in the square lattice. In (d), the blue lines show atomic vortices form the triangular lattice. While, the deviation of molecular vortices from the red lines indicates they form the amorphous state. In (e), atomic vortices form the triangular lattice and molecular vortices form the honeycomb lattices. Similarly, molecular vortices display the triangular lattice in (f).
The unit of length is $1.07\mu m$.

\textbf{Figure 3 The effect of Raman detuning on the lattice of atomic vortices with interstitial molecular vortices at the equilibrium state.}
Here the strength of interactions are $g_{m}=2g_{a}$ and $g_{am}=0.87g_{a}$. The rotation frequency is $\Omega=0.8\omega$. The upper plots [(a1)-(f1)] show the densities of atomic BEC and the lower plots [(a2)-(f2)] indicate the corresponding densities of molecular BEC. The value of Raman detuning varies.
(a) $\varepsilon=-2\hbar\omega$, (b) $\varepsilon=0\hbar\omega$, (c) $\varepsilon=2\hbar\omega$, (d) $\varepsilon=4\hbar\omega$, (e) $\varepsilon=7\hbar\omega$, and (f) $\varepsilon=10\hbar\omega$. (a1), (b1) and (c1) are the square lattice. (a2), (b2), (e1) and (f1) are the triangular lattice. Other plots show the amorphous state.
The unit of length is $1.07\mu m$.

\bigskip

\textbf{Figure 4 The number of vortices and particles.}
(a)-(d) show the number of atomic vortices $C_{a}$ and molecular vortices $C_{m}$ in atomic-molecular BECs of $^{87}$Rb with the detuning parameter $\varepsilon$ when the system reaches the equilibrium state.
 (a) $g_{am}=-0.87g_{a}$, (b) $g_{am}=0$, (c) $g_{am}=0.87g_{a}$ and (d) $g_{am}=2g_{a}$.
(e)-(h) indicate the corresponding particle number of atomic-molecular BECs of $^{87}$Rb, respectively.
 The rotation frequency is $\Omega=0.8\omega$, the strength of molecule-molecule interactions are $g_{m}=2g_{a}$ with the atom-atom scattering length $a_{a}=101.8a_{B}$, and the parameter $\chi$ is fixed to be $2\times10^{-3}$. The unit of detuning parameter is $\hbar\omega$.

\bigskip

\textbf{Figure 5 Phase diagram of rotating atomic-molecular BECs of $^{87}$Rb
when the system reaches the equilibrium state.} AMBEC(I) denotes the miscible mixture of atomic-molecular BECs, and AMBEC(II) is immiscible atomic-molecular BEC.
Furthermore, based on the phase diagram of atomic-molecular BECs, we further plot the phase diagram of atomic-molecular vortices when the atomic-molecular BECs of $^{87}$Rb reaches the equilibrium state. Then, the region of AMBEC(I) is divided into three parts: (1), (2), and (3). The overlapped atomic-molecular vortices, carbon-dioxide-type atomic-molecular vortices and atomic vortices with the interstitial molecular vortices occur in region (1), region (2) and region (3), respectively. In the green region, atomic and molecular vortices match fully with the rough ratio $1:2$.
 The parameters are $\Omega=0.8\omega$, $g_{m}=2g_{a}$ ($a_{a}=101.8a_{B}$), and $\chi=2\times10^{-3}$. The units of detuning parameter and $g_{am}$ are $\hbar\omega$ and $g_{a}$, respectively.

\bigskip

\textbf{Figure 6 The densities and phases of the atomic-molecular BECs of $^{87}$Rb under various rotating frequencies when the system reaches the equilibrium state.}
The rotating frequencies are indicated at the title of the subplots.
(a1)-(e1) are the densities of atomic BEC, (a2)-(e2) are the corresponding phases of atomic BEC, (a3)-(e3) are the densities of molecular BEC, and (a4)-(e4) are the corresponding phases of molecular BEC, respectively.
The critical rotating frequencies for inducing molecular vortex and atomic vortex are about $0.1\omega$ and $0.3\omega$, respectively.
The strength of atom-molecule interaction is $g_{am}=-0.87g_{a}$ with the atom-atom scattering length $a_{a}=101.8a_{B}$, molecule-molecule interactions is $g_{m}=2g_{a}$, the parameter $\chi$ is fixed to be $2\times10^{-3}$ and Raman detuning is $\varepsilon=0\hbar\omega$.
The unit of length is $1.07\mu m$.

\begin{figure}
\begin{center}
\epsfig{file=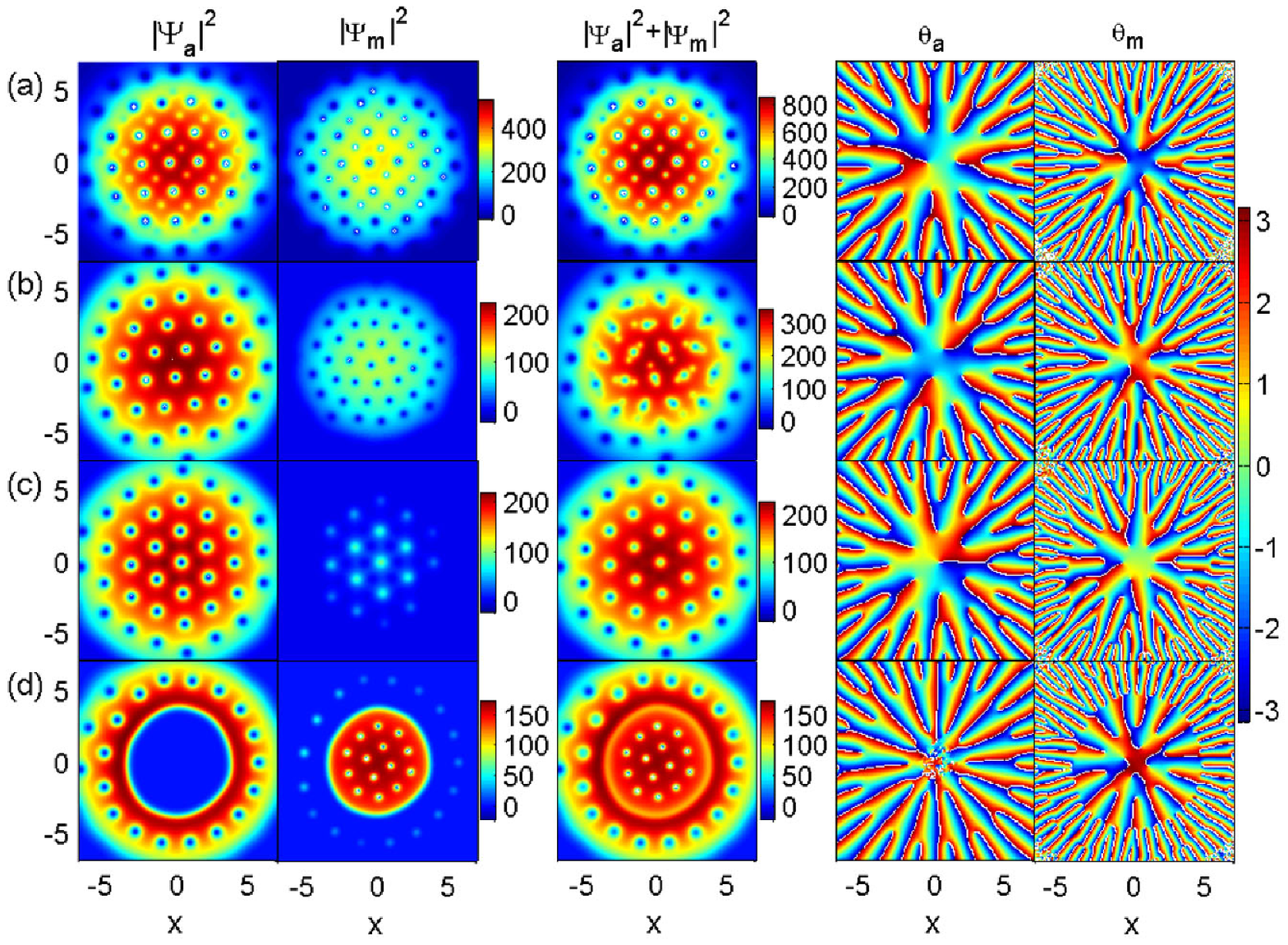,width=17cm}
\end{center}
\label{fig:TQPT}
\end{figure}
\bigskip

\newpage
\begin{figure}
\begin{center}
\epsfig{file=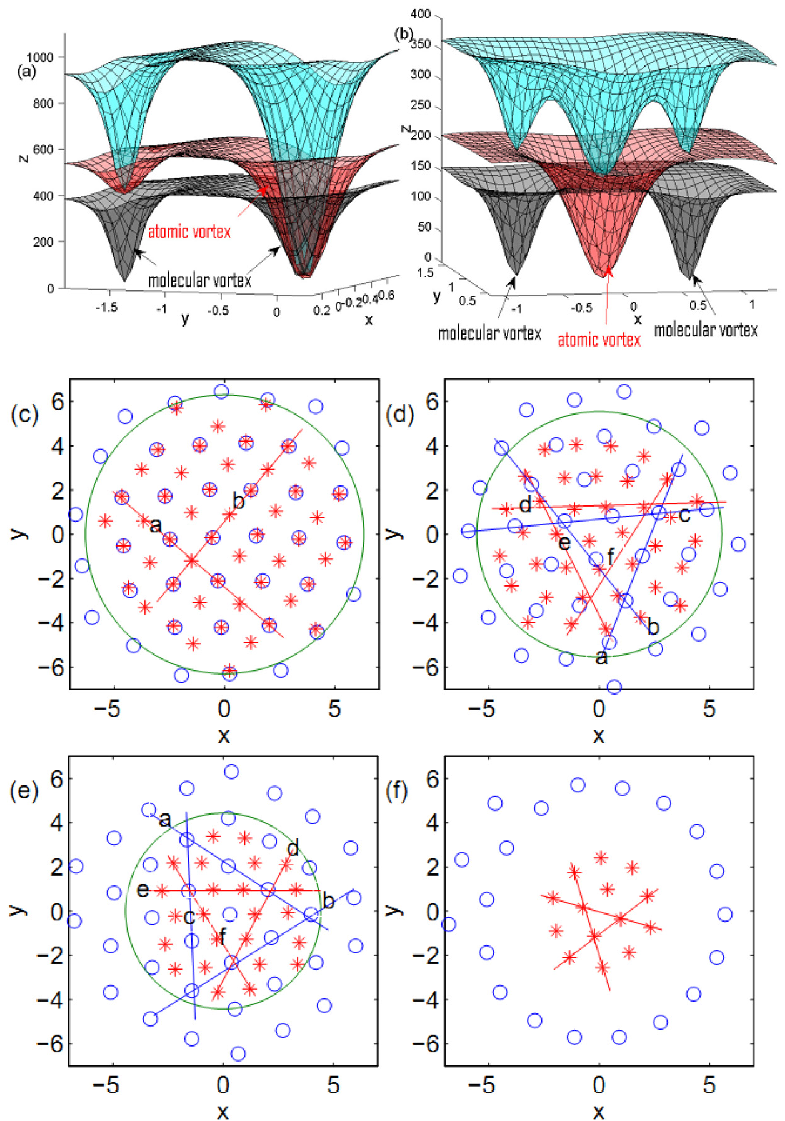,width=14cm}
\end{center}
\label{fig:FiniteT}
\end{figure}
\clearpage
\bigskip

\begin{figure}
\begin{center}
\epsfig{file=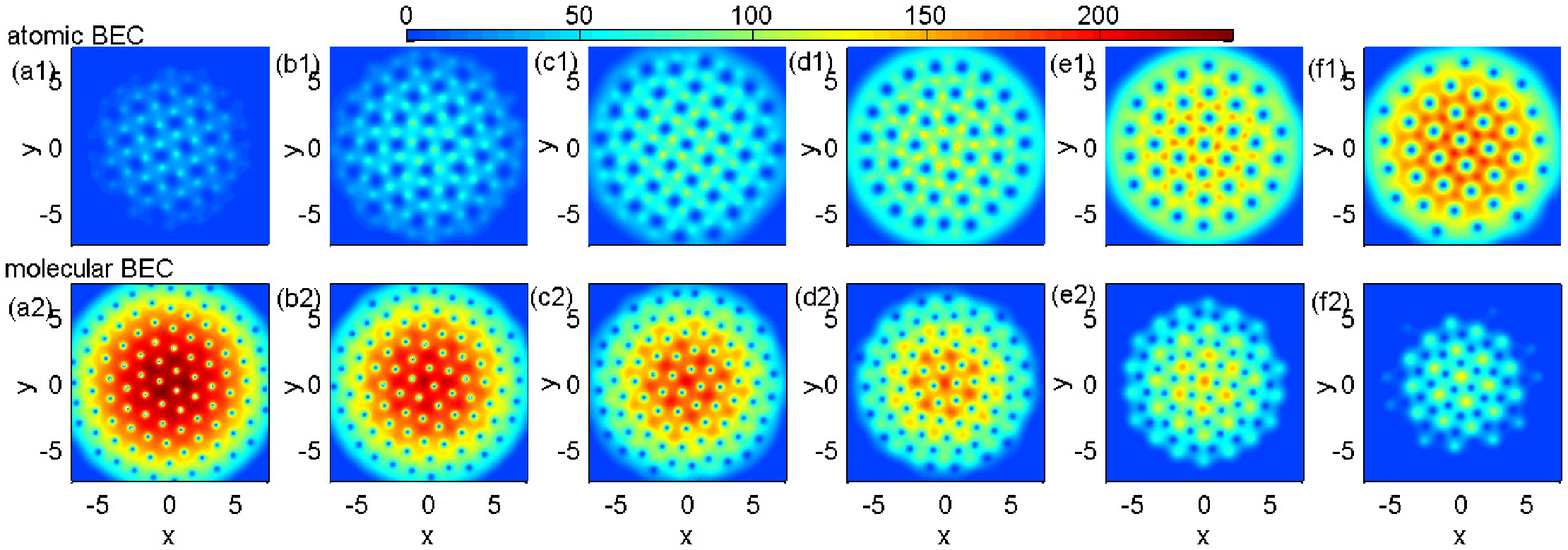,width=18cm, trim=0.5in 0.0in 0.0in 0in}
\end{center}
\label{fig:FiniteT}
\end{figure}
\clearpage
\bigskip

\begin{figure}
\begin{center}
\epsfig{file=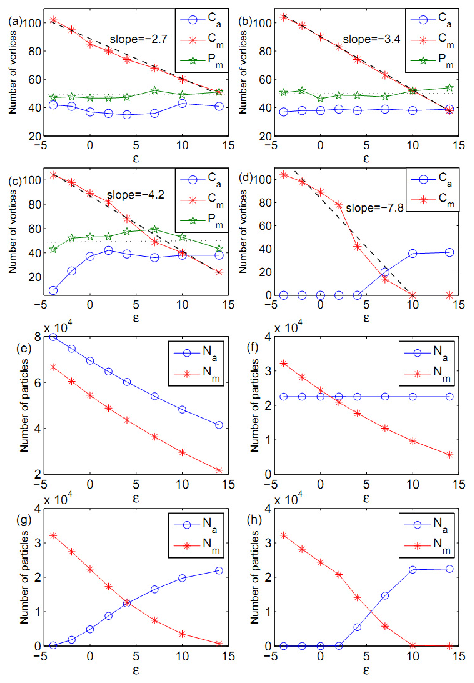,width=15cm}
\end{center}
\label{fig:FiniteT}
\end{figure}
\clearpage
\bigskip

\begin{figure}
\begin{center}
\epsfig{file=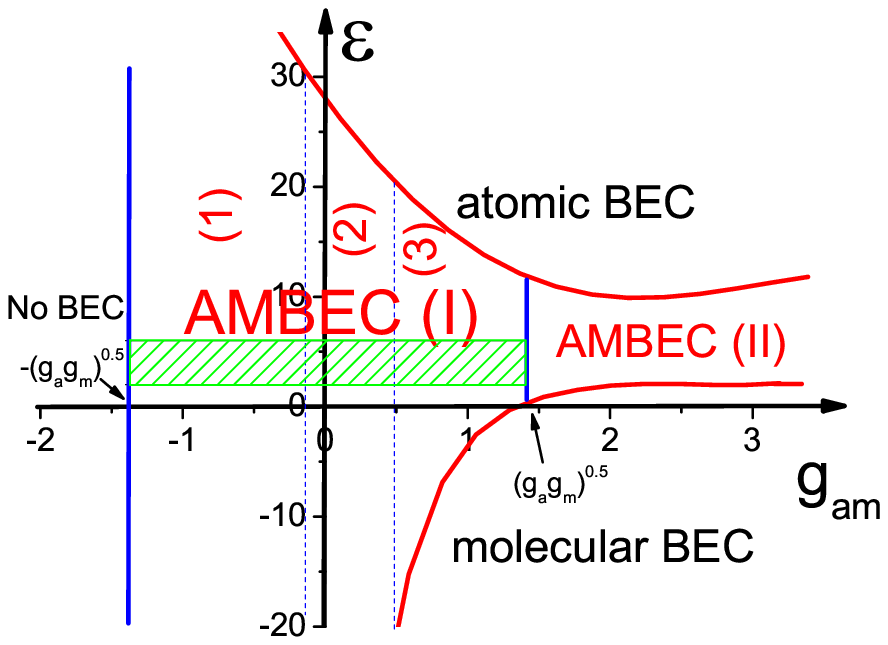,width=17cm}
\end{center}
\label{fig:ES}
\end{figure}

\newpage
\begin{figure}
\begin{center}
\epsfig{file=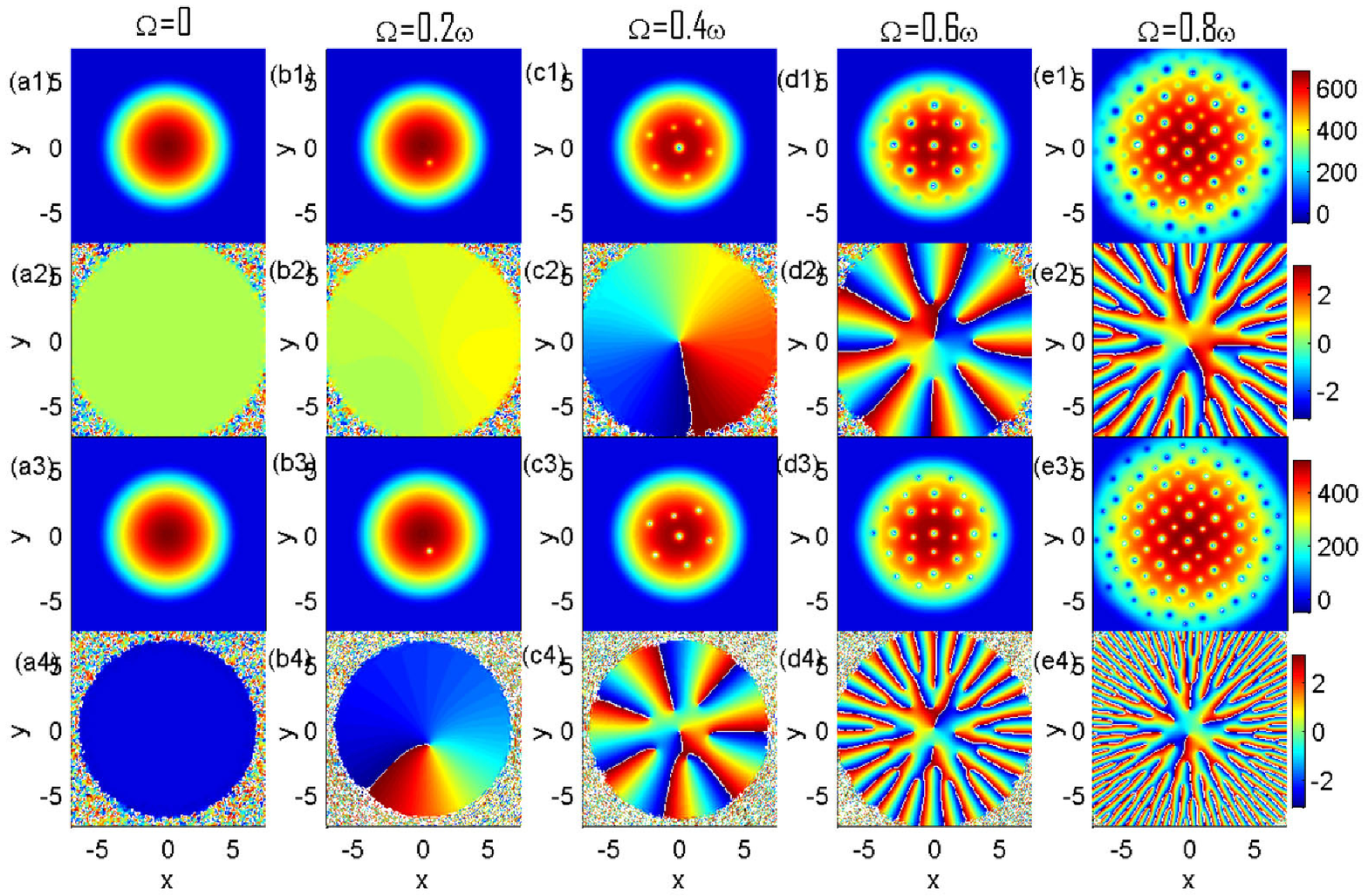,width=17cm}
\end{center}
\label{fig:HLPD}
\end{figure}
\bigskip

\begin{landscape}

\begin{table}
 \caption{\label{tab:test} A summary of the properties of vortices in the rotating atomic-molecular BECs of $^{87}$Rb when the system reaches the equilibrium state. The atomic-molecular vortices are composite in the matching region [inside the green circle in Figs. 2(c)-(e)]. }
\newsavebox{\tablebox}
\begin{lrbox}{\tablebox}

 \begin{tabular}{|p{.17\textwidth}|p{.33\textwidth}|p{.2\textwidth}|p{.22\textwidth}|p{.25\textwidth}|c|c|c|c|}\hline
Region (in Fig. 5) & Vortex state (in the matching region) & Lattice of atomic vortex & Lattice of molecular vortex & Vortex lattice out of the matching region &  \\\hline

 $(1)$    &   Overlapped atomic-molecular vortices with interstitial molecular vortices    &  square         &      square     &  triangular &  \\\hline

 $(2)$    &   carbon-dioxide-type atomic-molecular vortices      &  triangular         &      amorphous     &  triangular &  \\\hline

 $(3)$    &   atomic vortices with interstitial molecular vortices     &  square/ amorphous/ triangular       &   triangular/ amorphous /honeycomb    &  triangular &  \\\hline

 AMBEC(II)    &   separated atomic vortices and molecular vortices   &  triangular          &      triangular      &  No & \\\hline

 atomic BEC    &   pure atomic vortices      &  triangular          &      No     &  No &  \\\hline

 molecular BEC    &   pure molecular vortices        &  No        &      triangular     &  No &  \\\hline

 \end{tabular}
\end{lrbox}
\rotatebox[origin=c]{0}{\usebox{\tablebox}}
\end{table}

\end{landscape}


\begin{thebibliography}{99}

\bibitem{Madison}
Madison, K. W., Chevy, F., Wohlleben, W. \& Dalibard, J.
Vortex formation in a stirred Bose-Einstein condensate.
 \textit{Phys. Rev. Lett.} \textbf{84}, 806 (2000).



\bibitem{AboShaeer}
Abo-Shaeer, J. R., Raman, C., Vogels, J. M. \& Ketterle, W.
Observation of vortex lattices in Bose-Einstein condensates.
 \textit{Science} \textbf{292}, 476 (2001).

\bibitem{Engels}
Engels, P., Coddington, I., Haljan, P. C. \& Cornell, E. A.
Nonequilibrium effects of anisotropic compression applied to vortex lattices in Bose-Einstein condensates.
\textit{Phys. Rev. Lett.} \textbf{89}, 100403 (2002).

\bibitem{Cooper} Cooper, N. R., Rezayi, E. H. \& Simon, S.H.,
Vortex lattices in rotating atomic Bose gases with dipolar interactions.
\textit{Phys. Rev. Lett.} \textbf{95}, 200402 (2005).


\bibitem{Kasamatsu}
Kasamatsu, K., Tsubota, M. \& Ueda, M.
Vortex phase diagram in rotating two-component Bose-Einstein condensates.
\textit{Phys. Rev. Lett.} \textbf{91}, 150406 (2003).

\bibitem{Kuopanportti}
Kuopanportti, P., Huhtam\"{a}ki, Jukka, A. M. \& M\"{o}tt\"{o}nen, M.,
Exotic vortex lattices in two-species Bose-Einstein condensates.
\emph{Phys. Rev. A} \textbf{85}, 043613 (2012).


\bibitem{Wucj}
Zhou, X. F., Zhou, J. \& Wu, C. J.
Vortex structures of rotating spin-orbit-coupled Bose-Einstein condensates.
\emph{Phys. Rev. A} \textbf{84}, 063624 (2011).

\bibitem{Malomeda}
Mihalache, D., Mazilu, D., Malomed, B. A. \& Lederer, F.
Vortex stability in nearly-two-dimensional Bose-Einstein condensates with attraction.
\emph{Phys. Rev. A} \textbf{73}, 043615 (2006).

\bibitem{Malomedb}
Wu, L. \emph{et al.} 
Exact solutions of the Gross-Pitaevskii equation for stable vortex modes in two-dimensional Bose-Einstein condensates.
\emph{Phys. Rev. A} \textbf{81}, 061805(R) (2010).



\bibitem{Jiac}
Ji, A. C., Liu, W. M., Song, J. L. \& Zhou, F.
Dynamical creation of fractionalized vortices and vortex lattices.
\textit{Phys. Rev. Lett.} \textbf{101}, 010402 (2008).

\bibitem{Wangds}
Wang, D. S., Song, S. W., Xiong, B. \& Liu, W. M.
Vortex states in a rotating Bose-Einstein condensate with spatiotemporally modulated interaction.
\textit{Phys. Rev. A} \textbf{84}, 053607 (2011).

\bibitem{Hanwei}
Han, W., Zhang, S. Y., Jin, J. J. \& Liu, W. M.
Half-vortex sheets and domain-wall trains of rotating two-component Bose-Einstein condensates in spin-dependent optical lattices.
\textit{Phys. Rev. A} \textbf{85}, 043626 (2012).




\bibitem{liuDPGPE} Liu, C. F., Yu, Y. M., Gou, S. C. \& Liu, W. M.
Vortex chain in anisotropic spin-orbit-coupled spin-1 Bose-Einstein condensates.
\textit{Phys. Rev. A} \textbf{87}, 063630 (2013).

\bibitem{CFLiu} Liu C. F. \& Liu, W. M.
Spin-orbit-coupling-induced half-skyrmion excitations in rotating and rapidly quenched spin-1 Bose-Einstein condensates.
\textit{Phys. Rev. A} \textbf{86}, 033602 (2012).

\bibitem{liucf} Liu, C. F. \emph{et al.} 
Circular-hyperbolic skyrmion in rotating pseudo-spin-1/2 Bose-Einstein condensates with spin-orbit coupling.
\textit{Phys. Rev. A} \textbf{86}, 053616 (2012).


\bibitem{Su} Su, S. W. \emph{et al.} 
Spontaneous crystallization of skyrmions and fractional vortices in fast-rotating and rapidly quenched spin-1 Bose-Einstein condensates.
\textit{Phys. Rev. A} \textbf{84}, 023601 (2011).









\bibitem{Woo} Woo, S. J., Park, Q. H. \& Bigelow, N. P.
Phases of atom-molecule vortex matter.
\textit{Phys. Rev. Lett.} \textbf{100}, 120403 (2008).

\bibitem{Alexander}
Alexander, T. J., Ostrovskaya, E. A., Kivshar, Y. S. \& Julienne, P. S.
Vortices in atomic-molecular Bose-Einstein condensates.
\textit{J. Opt. B: Quantum Semiclass. Opt.} \textbf{4}, S33 (2002).





\bibitem{Timmermans}
Timmermans, E., Tommasini, P., C\^{o}t\'{e}, R.,  Hussein, M. \&  Kerman, A.
Raried liquid properties of hybrid atomic-molecular Bose-Einstein condensates.
\textit{Phys. Rev. Lett.} \textbf{83}, 2691 (1999).

\bibitem{Donley}
Donley, E. A., Claussen, N. R., Thompson, S. T. \& Weiman, D. E.
Atom-molecule coherence in a Bose-Einstein condensate.
\textit{Nature (London)} \textbf{417}, 529 (2002).



\bibitem{Cusack}
Cusack, B. J., Alexander, T. J., Ostrovskaya E. A. \& Kivshar, Y. S.
Existence and stability of coupled atomic-molecular Bose-Einstein condensates.
\textit{Phys. Rev. A} \textbf{65}, 013609 (2001).


\bibitem{Basu} Basu, S. \& Mueller, E. J.
Stability of bosonic atomic and molecular condensates near a Feshbach resonance.
\textit{Phys. Rev. A} \textbf{78}, 053603 (2008).


\bibitem{Xu}
Xu, X. Q., Lu, L. H. \& Li, Y. Q.
Phase separation in atom-molecule mixtures near a Feshbach resonance.
\textit{Phys. Rev. A} \textbf{79}, 043604 (2009).

\bibitem{Zhou}
Zhou, L., Qian, J., Pu, H., Zhang, W. \& Ling, H. Y.
Phase separation in a two-species atomic Bose-Einstein condensate with an interspecies Feshbach resonance.
\textit{Phys. Rev. A} \textbf{78}, 053612 (2008).


\bibitem{Ling}
Ling, H. Y., Pu, H. \& Seaman, B.
Creating a stable molecular condensate using a generalized Raman adiabatic passage scheme.
\textit{Phys. Rev. Lett.} \textbf{93}, 250403 (2004).


\bibitem{Linghy}
Ling, H. Y., Maenner, P., Zhang, W. P. \& Pu, H.
Adiabatic theorem for a condensate system in an atom-molecule dark state.
\textit{Phys. Rev. A} \textbf{75}, 033615 (2007).



\bibitem{Wynar}
Wynar, R., Freeland, R. S., Han, D. J., Ryu, C. \& Heinzen, D. J.
Molecules in a Bose-Einstein condensate.
\textit{Science} \textbf{287}, 1016 (2000).



\bibitem{McKenzie} McKenzie, C. \emph{et al.} 
Photoassociation of sodium in a Bose-Einstein condensate.
\textit{Phys. Rev. Lett.} \textbf{88}, 120403 (2002).

\bibitem{Abdullaev}
Abdullaev, F. Kh. \& Konotop, V. V.
Intrinsic localized modes in arrays of atomic-molecular Bose-Einstein condensates.
\textit{Phys. Rev. A} \textbf{68}, 013605 (2003).


\bibitem{Cruz}
Cruz, H. A. \& Konotop, V. V.
Inhomogeneous dark states of atomic-molecular Bose-Einstein condensates in trapping potentials.
\textit{Phys. Rev. A} \textbf{83}, 033603 (2011).



\bibitem{Heinzen} Heinzen, D. J., Wynar, R., Drummond, P. D. \& Kheruntsyan, K. V.
Superchemistry: dynamics of coupled atomic and molecular Bose-Einstein condensates.
\textit{Phys. Rev. Lett.} \textbf{84}, 5029 (2000).


\bibitem{Drummond} Drummond, P. D., Kheruntsyan, K. V., Heinzen D. J. \& Wynar, R. H.
Stimulated Raman adiabatic passage from an atomic to a molecular Bose-Einstein condensate.
\textit{Phys. Rev. A} \textbf{65}, 063619 (2002).


\bibitem{Gupta} Gupta, M. \& Dastidar, K. R.
Control of the dynamics of coupled atomic-molecular Bose-Einstein condensates: Modified Gross-Pitaevskii approach.
\textit{Phys. Rev. A} \textbf{80}, 043618 (2009).



\bibitem{Drummo} Drummond, P. D., Kheruntsyan, K. V. \& He, H.
Coherent Molecular Solitons in Bose-Einstein Condensates.
\textit{Phys. Rev. Lett.} \textbf{81}, 3055 (1998).

\bibitem{Hope} Hope, J. J., \& Olsen, M. K.
Quantum Superchemistry: Dynamical Quantum Effects in Coupled Atomic and Molecular Bose-Einstein Condensates.
\textit{Phys. Rev. Lett.} \textbf{86}, 3220 (2001).



\bibitem{Herzog} Herzog, C. \& Olshanii, M.
Trapped Bose gas: The canonical versus grand canonical statistics.
\textit{Phys. Rev. A} \textbf{55}, 3254 (1997).


\bibitem{Kocharovsky} Kocharovsky, V. V., Scully, M. O., Zhu, S. Y. \& Suhail Zubairy, M.
Condensation of N bosons. II. Nonequilibrium analysis of an ideal Bose gas and the laser phase-transition analogy.
\textit{Phys. Rev. A} \textbf{61}, 023609 (2000).

\bibitem{Cockburn} Cockburn, S. P., Negretti, A., Proukakis, N. P. \& Henkel, C.
Comparison between microscopic methods for finite-temperature Bose gases.
\textit{Phys. Rev. A} \textbf{83}, 043619 (2011).




\bibitem{Lin} Lin, S. Z. \& Bulaevskii, L. N.
Dissociation transition of a composite lattice of magnetic vortices in the flux-flow regime of two-band superconductors.
\textit{Phys. Rev. Lett.} \textbf{110}, 087003 (2013).

%



\bibitem{Tikhonenkov} Tikhonenkov, I. \& Vardi, A.
Atom-molecule dephasing in an SU(1,1) interferometer based on the stimulated dissociation of a molecular Bose-Einstein condensate.
\textit{Phys. Rev. A} \textbf{80}, 051604(R) (2009).

\bibitem{Fischer} Fischer U. R. and Baym G.,
Vortex States of Rapidly Rotating Dilute Bose-Einstein Condensates.
\textit{Phys. Rev. Lett.} \textbf{90} 140402 (2003).

\bibitem{Rooney} Rooney, S. J., Bradley, A. S. \& Blakie, P. B.
Decay of a quantum vortex: Test of nonequilibrium theories for warm Bose-Einstein condensates.
\textit{Phys. Rev. A} \textbf{81}, 023630 (2010).




\bibitem{Bradley} Bradley, A. S., Gardiner, C. W. \& Davis, M. J.
Bose-Einstein condensation from a rotating thermal cloud: Vortex nucleation and lattice formation.
\textit{Phys. Rev. A} \textbf{77}, 033616 (2008).


















\end{thebibliography}
\end{document}